\documentclass[12pt]{article}

\usepackage{amssymb}
\usepackage{bbm}
\usepackage{epsfig}
\usepackage{array}
\usepackage{float}
\usepackage{dsfont}
\usepackage{amstext}
\usepackage{amsmath}
\usepackage{psfrag}
\usepackage{a4}
\usepackage{a4wide}

\usepackage{hyperref}

\usepackage{feynmp-auto}
\def\be{\begin{equation}}
\def\ee{\end{equation}}
\def\bea{\begin{eqnarray}}
\def\eea{\end{eqnarray}}

\begin{document}

\begin{center}
\baselineskip 20pt 
{\Large\bf New Inflation in Supersymmetric SU(5) and \\
Flipped SU(5) GUT Models}

\vspace{1cm}

{\large 
Mansoor Ur Rehman$^{}$\footnote{Email: \texttt{\href{mailto:mansoor@qau.edu.pk}{mansoor@qau.edu.pk}}},
Mian Muhammad Azeem Abid$^{}$\footnote{Email:\texttt{\href{mailto:azeem_92@live.com}{azeem\_92@live.com}}} and Amna Ejaz$^{}$\footnote{Email: \texttt{\href{mailto:ejazamna33@gmail.com}{ejazamna33@gmail.com}}} 
} 
\vspace{.5cm}

{\baselineskip 20pt \it
$^a$Department of Physics,  \\
Quaid-i-Azam University, Islamabad 45320, Pakistan 
\vspace{2mm} }

\vspace{1cm}

\end{center}

\begin{abstract}
	We study the feasibility of realizing supersymmetric new inflation model, introduced by Senoguz and Shafi in \cite{Senoguz:2004ky}, for $SU(5)$ and flipped $SU(5)$ models of grand unified theories (GUTs). This realization requires an additional $U(1)_R \times Z_{n}$ symmetry for its successful implementation. The standard model (SM) gauge singlet scalar components of $24_H$ and $10_H$ GUT Higgs superfields are respectively employed to realize successful inflation in $SU(5)$ and flipped $SU(5)$ models.  The predictions of the various inflationary observables lie within the recent Planck bounds on the scalar spectral index, $n_s$, for $n \geq 5$ in $SU(5)$ model and for $n \geq 6$ in flipped $SU(5)$ model.  In particular, the tensor to scalar ratio $r$ and the running of spectral index $d n_s/ d\ln k$ are negligibly small and lie in the range, $10^{-12} \lesssim r \lesssim 10^{-8}$ and $10^{-9} \lesssim dn_s/d\ln k \lesssim 10^{-3}$, for realistic values of $n$. In numerical estimation of the various predictions, we fix the gauge symmetry breaking scale, $M$, around $2 \times 10^{16}$ GeV. The issue of gauge coupling unification in $R$-symmetric $SU(5)$ is evaded by adding vectorlike families with mass splitting within their multiplets. The dilution of monopoles beyond the observable limit is naturally achieved  in the breaking of $SU(5)$ gauge symmetry during inflation. A realistic scenario of reheating with non-thermal leptogenesis is employed for both models. The predicted range of reheat temperature within Planck bounds, $3 \times 10^{7}\text{ GeV }\lesssim T_r \lesssim 2 \times 10^{9}$ GeV, is safe from the gravitino problem for the gravitino mass, $m_{3/2} \gtrsim 10$ TeV. Finally, the $U(1)_R \times Z_{n}$ symmetry is also observed to play a crucial role in suppressing the various fast proton decay operators. 
\end{abstract}

%
\section{\large{\bf Introduction}}%
The three early models of inflation namely, new \cite{Linde:1981mu}, chaotic \cite{Linde:1983gd} and hybrid inflation \cite{Linde:1993cn}, have gained the most attention among the various proposed models of inflation. As compared to chaotic, new and hybrid are regarded as the small field inflation models and their supersymmetric realization has also been a subject of great interest. Specifically, the model of supersymmetric hybrid inflation \cite{Dvali:1994ms,Copeland:1994vg,Linde:1997sj,Chamseddine:1982jx,Senoguz:2004vu,Senoguz:2003zw} has developed an intimate connection of inflation with the models of grand unified theories (GUTs). Further, to avoid GUT monopole problem, the smooth and shifted extensions of supersymmetric hybrid inflation were proposed \cite{Lazarides:1995vr,Jeannerot:2000sv}. 
Particularly, a $U(1)_R \times Z_2$ symmetry was employed in smooth hybrid inflation. The first supersymmetric model of new inflation was proposed in \cite{Izawa:1996dv,Izawa:1997df} and 
in order to explain its initial conditions dynamically a model of smooth pre-inflation, based on a generalized $U(1)_R\times Z_n$ symmetry, was considered in \cite{Yamaguchi:2004tn}. Later, Senoguz and Shafi (SS) utilized this generalized smooth hybrid model to realize a supersymmetric model of new inflation \cite{Senoguz:2004ky}. In this SS model of new inflation, the scalar component of GUT Higgs superfield, charged under the $Z_n$ symmetry, acts as an inflaton whereas the gauge singlet superfield with non-zero $R$ charge is stabilized at zero. In addition, a successful realization of non-thermal leptogenesis was presented in this model to explain the observed baryon asymmetry. For other relevant models of new inflation, see \cite{Antusch:2008gw,Antusch:2013eca}.

In present article we discuss the feasibility of realizing SS new inflation model in $SU(5)$ GUT \cite{Georgi:1972cj} and its flipped version, i.e., $FSU(5) \equiv SU(5) \times U(1)$ GUT \cite{Barr:1981qv}. Some of the distinguishing features of these models are the solution of doublet-triplet problem, the gauge coupling unification and the appropriate proton decay suppression. As we will discuss in detail, the $U(1)_R \times Z_n$ symmetry plays a crucial role in circumventing the fast proton decay problem. A fine-tuned solution is assumed for the doublet-triplet problem in $SU(5)$ model whereas this problem is naturally resolved in $FSU(5)$ via missing partner mechanism \cite{Antoniadis:1987dx}. The additional $R$-symmetry ruins the successful gauge coupling unification feature of a supersymmetric $SU(5)$ model and inevitably leads us beyond the minimal setup of the model. The gauge coupling unification feature, however, is not predicted in $FSU(5)$ model. The problem of light neutrino masses with successful reheating and leptogenesis is naturally settled in $SU(5)$ model whereas a little extra effort is required to realize these features in the $R$-symmetric $FSU(5)$ model considered here. The monopole problem appears in $SU(5)$ model with the prediction of electric charge quantization. However, the breaking of $SU(5)$ symmetry during new inflation automatically takes care of this problem. The observed mass hierarchy of charged fermions is easily accommodated in $FSU(5)$ whereas in $SU(5)$ we need to go beyond the minimal model. Finally, the predictions of the various inflationary parameters are found compatible with the recent Planck data based on a base $\Lambda$CDM model \cite{Ade:2015lrj}.

The presentation of the paper is as follows. In the second section we introduce $SU(5)$ model in the framework of SS new inflation model and derive the form of its inflationary potential. The SS new inflation model based on flipped $SU(5)$ GUT is introduced in the third section along with the derivation of inflationary potential. The fourth section is mainly devoted to the presentation and the discussion of the numerical predictions of the various important inflationary parameters. These predictions rely on reheating with successful non-thermal leptogenesis as is discussed in section-5. The role of the underlying symmetries in eliminating the fast proton decay operators is highlighted in section-6. The issue of gauge coupling unification in $R$-symmetric $SU(5)$ model and its fine-tuned solution is presented in section-7. Finally, we conclude our results and discussion in section-8.


\section{\large{\bf SUSY SU(5) hybrid inflation}}
In supersymmetric $SU(5)$ model, the minimal supersymmetric standard model (MSSM) matter content with the right handed neutrino superfield reside respectively in $\overline{5}+10$ and $1$ dimensional representations as, 
\bea
\overline{5}_i &=& D^c_i(\overline{3},1,1/3) + L_i(1,2,-1/2),  \nonumber \\
10_i &=& Q_i(3,2,1/6)+ U^c_i (\bar{3},1,-2/3) + E^c_i (1,1,+1),  \nonumber \\
1_i  &=& N^c_i(1,1,0),   \label{rep1}
\eea
where $N^c$ is the right handed neutrino superfield. The Higgs superfield responsible for the $SU(5)$ GUT breaking into MSSM gauge group belongs to the adjoint representation $24_H$ while the electroweak doublet Higgs pair, along with a color triplet Higgs pair, resides in the fundamental representations $5_{h}$, $\overline{5}_{h}$. With the addition of a gauge singlet superfield, $S$, the superpotential for new inflation can be written as,
\bea
		W_{SU(5)} &=& S\bigg(-\mu^{2}+\frac{\text{Tr}(24_H^{m})}{\Lambda^{m-2}} + \cdots \bigg)  \nonumber \\
		&+& \delta \, \overline{5}_h 5_h 
		+ \beta \dfrac{\text{Tr}(24_H^{m-1})}{\Lambda^{m-1}} \bar{5}_h 24_H 5_h + \cdots \nonumber \\
		&+& y_{ij}^{(u)}10_{i}10_{j}5_h+y_{ij}^{(d,e)}10_{i}\overline{5}_{j}\overline{5}_h+y_{ij}^{(\nu)}1_{i}\overline{5}_{j}5_h+m_{\nu_{ij}}1_{i}1_{j},  \label{superpot1}
\eea
where, $\mu$ is superheavy mass, $\Lambda$ is the cut-off scale and $m$ is an integer. Additional $ U(1)_{R} \times Z_{n}$ symmetry is also imposed on the superpotential. Under $ Z_{n}$ symmetry, only the superfield $24_H$ has unit charge with $m=n$ while under $U(1)_{R}$ symmetry the charges $q_R$ of the various superfields are assigned as,
\begin{equation}
	q_R  (S,24_H,5_h,\overline{5}_h,10_i,\overline{5}_i,1_i) = (1,0,2/5,3/5,3/10,1/10,1/2). \label{Rsu5}
\end{equation}
The dots in the above superpotential indicates additional possible gauge-invariant terms for $m\geq 4$. In the discussion below we have effectively combined these extra terms into the ones already indicated in the above superpotential.

In the above superpotential only the linear terms in $S$ are relevant for new inflation where the scalar SM gauge-singlet component of $24_H$ plays the role of inflaton. As $SU(5)$ gauge symmetry is broken during inflation, the monopoles produced are inflated away while diluting the monopole density below the observational limits. The same problem also arises in the standard hybrid inflation model based on the above superpotential with $m=2$. A shifted or a smooth version of SUSY hybrid inflation model is usually employed to avoid this problem. These variants of SUSY hybrid inflation model are discussed in \cite{Khalil:2010cp,Rehman:2014rpa} for $SU(5)$ GUT. 

The second-line terms in Eq.~(\ref{superpot1}) are involved in the solution of doublet-triplet problem. To make the electroweak Higgs-doublet pair light and the color Higgs-triplet pair heavy, a fine-tuned solution is assumed with $\delta \sim \beta \langle 24_H^{m} \rangle / \Lambda^{m-1}$. In the third line of Eq.~(\ref{superpot1}), the Dirac mass terms of quarks and leptons are generated by the Yukawa couplings $y_{ij}^{(u)}$, $y_{ij}^{(d,e)}$ and $y_{ij}^{(\nu)}$ whereas, $m_{\nu_{ij}}$, represents the right handed neutrino mass matrix. The last two terms are responsible for the generation of light  neutrino masses through the seesaw mechanism. Finally, an additional $Z_2$ matter parity is imposed to realize the lightest supersymmetric particle (LSP) a viable candidate for cold dark matter. A relevant discussion of fast proton decay in $SU(5)$ GUT is also presented in section-6.

To discuss the realization of new inflation model in $SU(5)$ we first consider the following relevant part of the superpotential from Eq.~(\ref{superpot1}), 
\begin{equation}
	W_{SU(5)} \supset S\bigg(-\mu^{2}+\frac{\text{Tr}(\Phi^{m})}{\Lambda^{m-2}}\bigg),
\end{equation}
where the superfield $\Phi$ represents the SM gauge-singlet component of $24_H$, 
\be
\Phi = \frac{\phi}{\sqrt{15}} \text{ diag}(-1,-1,-1,3/2,3/2).
\ee
The global SUSY vacuum lies at $ \langle S\rangle=0 $ and $ \langle \text{Tr}(\Phi)^{m} \rangle \equiv v^m = \mu^{2} \Lambda^{m-2}$, where $v $ is related to the symmetry breaking scale $ \langle \phi \rangle \equiv M$ as,
\begin{equation}
	M = \frac{\sqrt{15} \, v}{\sqrt[m]{3(-1)^m + 2(3/2)^m}} \simeq (2.49 - 2.58)\,v,
\end{equation}
for $m = 3$ to infinity. The supergravity (SUGRA) scalar potential is given by
\begin{equation}
	V = {\rm e}^K\left[\left(\frac{\partial^2 K}{\partial z_i\partial z^*_j}\right)^{-1}D_{z_i}W D_{z^*_j}W^*-3|W|^2\right] + V_D\,,  \label{potformula}
\end{equation}
with
\begin{equation}
	D_{z_i}W=\frac{\partial W}{\partial z_i}+\frac{\partial K}{\partial z_i}W\, \text{ and } z_i=\{S,\phi,...\}.
\end{equation}
The K\"ahler potential $K$ with non-minimal terms expanded up to second order in $ 1/m_P $ can be written  as,
\bea
K  &=& \bigg(1 + k_1 w_1 + \cdots \bigg)\left( \Lambda^2 + |S|^{2}+\text{Tr}|\Phi|^{2}+\kappa_{1}\dfrac{\text{Tr}|\Phi|^{3}}{m_{P}} \right. + \cdots \\
&+& \left. \kappa_{S\Phi} \dfrac{|S|^{2}\text{Tr}|\Phi|^{2}}{m_{P}^{2}}+\kappa_{S} \dfrac{|S|^{4}}{4m_{P}^{2}}+\kappa_{\Phi} \dfrac{(\text{Tr}|\Phi|^{2})^{2}}{4m_{P}^{2}}+ \kappa_{2} \dfrac{\text{Tr}|\Phi|^{4}}{4m_{P}^{2}} + \cdots \right) + \cdots ,
\eea
where,
\be
w_1 = \frac{\text{Tr}(\Phi^{m})}{\Lambda^{m}}+\text{h.c.}.
\ee
The relevant part of the D-term scalar potential, 
\begin{equation}
	V_{D} \supset g^{2} \bigg(f^{ijk}24_{H_{j}} 24_{H_{k}}^{\dagger} \bigg) 
	\bigg( f^{ilm}24_{H_{l}} 24_{H_{m}}^{\dagger} \bigg),
\end{equation}
vanishes for the component chosen along the 24 direction, since $ f^{i,24,24}=0 $. Therefore, this direction also becomes the D-flat direction with  other gauge non-singlet fields stabilized at zero.

Using the above expressions along with the D-flat direction, the scalar potential for $(|\phi|$, $|S|)\ll m_P$ becomes
\begin{equation}
	V \simeq  \mu^{4} \bigg| 1 - \bigg( \frac{\phi}{M} \bigg)^{m} \bigg|^{2} 
	+ \mu^{4} \bigg| \frac{m \phi^{m-1}}{M^{m}} \bigg|^{2} | S |^{2}
	+  \mu^{4}\bigg[\bigg(\dfrac{1-\kappa_{S\Phi}}{(m_{P}/v)^2}\bigg) \bigg| \frac{\phi}{M} \bigg|^2-\bigg(\dfrac{\kappa_{S}}{m_{P}^{2}}\bigg)|S|^{2}\bigg].
\end{equation} 
Here we have used the same notation for superfields and their scalar components. In order to eliminate the $S$ field we assume $\kappa_{S}< -1/3 $. The $S$ field thus acquires a mass larger than the Hubble scale, $H \simeq \mu^2/\sqrt{3}\,m_P$, and is stabilized to zero instantly \cite{Senoguz:2004ky}. This leaves us with a potential of a complex field, $\phi = |\phi|e^{i\arg(\phi)}$. We further assume appropriate initial condition for the phase, $\arg(\phi)$, to remain stabilized at zero during inflation. The dependence of inflationary dynamics on this phase factor with general initial conditions can be seen in \cite{Nolde:2013bha}. We define, $ \gamma \equiv \frac{\kappa_{S\Phi}-1}{(M/v)^2}\geq0 $, and set the symmetry breaking scale, $M$, equal to the GUT scale $M_{GUT}$, i.e. $ M = M_{GUT}\simeq 2\times10^{16}\text{ GeV} $.  
To obtain a canonically normalized field we replace $\phi \rightarrow \phi/\sqrt{2}$. Thus the scalar potential for new inflation with $\phi \ll M$ takes the following form,
\begin{equation}
	V_{SU(5)}\simeq\mu^{4}\bigg(1 - \frac{\gamma}{2}\frac{\phi^{2}}{m_{P}^{2}}- 2 \bigg(\frac{\phi}{\sqrt{2}M}\bigg)^{m} + \bigg(\frac{\phi}{\sqrt{2}M}\bigg)^{2m} \bigg),  \label{vsu5}
\end{equation} 
where the field $\phi$ now represents the canonically normalized real scalar field. The presentation of results and discussion for this model is delayed until we first derive a similar form of the potential for the new inflation model based on flipped $SU(5)$ GUT.

\section{\large{\bf SUSY FSU(5) hybrid inflation}}%
In supersymmetric $Flipped$ $SU(5)\equiv FSU(5)=SU(5)\times U(1)$ model, the MSSM matter content with the right handed neutrino superfield resides in the following representations, 
\bea
10_i &=& (10,1)_i = Q_i(3,2,1/6)+ D^c_i(\overline{3},1,1/3) + N^c_i(1,1,0),  \nonumber \\
\overline{5}_i &=& (\overline{5},-3)_i = U^c_i (\bar{3},1,-2/3) + L_i(1,2,-1/2),    \nonumber \\
1_i  &=& (1,5)_i = E^c_i (1,1,+1).   \label{rep2}
\eea
Note that the embedding of the matter superfields $U^c \leftrightarrow D^c$ and $E^c \leftrightarrow N^c$ in the $FSU(5)$ representations shown above are flipped as compared to the corresponding $SU(5)$ superfields embedding in Eq.~(\ref{rep1}), and this provides the reason why former is called the flipped $SU(5)$ model \cite{Barr:1981qv}. The Higgs superfield responsible for the $FSU(5)$ GUT breaking into MSSM belongs to the $10_H$, $\overline{10}_H$ pair containing the SM gauge-singlet components $\Phi$ and $\overline{\Phi}$ respectively. Similar to $SU(5)$ model, the two Higgs pairs of electroweak doublet and a color triplet reside in the fundamental representations $5_{h}$, $\overline{5}_{h}$. 
The charge assignments of the various superfields  under $U(1)_R$ and $Z_n$ symmetries are shown in the Table-I. An extra $Z_2$ matter parity, as defined in Table-I, is also required to make the lightest SUSY particle the dark matter candidate.
\begin{table}
\begin{center}
\begin{tabular}{|c||c|cc|cc|ccc|}
\hline
& ~$S$& ${\bf 10}_H$ & ${\bf \overline{10}}_H$ & ${\bf
5}_h$ & ${\bf\overline{5}}_h$ & ${\bf 10}_i$ &
${\bf\overline{5}}_i$ & ${\bf 1}_i$
\\ \hline\hline
$U(1)_R$ & $1$ & $0$ & $0$ & $1$ & $1$ & $0$ & $0$ & $0$
\\
$Z_n$ & $0$ & $1$ & $1$ & $n-2$ & $n-2$ & $1$ & $1$ & $1$
\\
$Z_2$ & $+$ & $+$ & $+$ & $+$ & $+$ & $-$ & $-$ & $-$
\\ \hline
\end{tabular}
\end{center}
\caption{The superfield content of $FSU(5)$ with charge assignments under $U(1)_R \times Z_n$ and $Z_2$ matter parity.}
\end{table}

The $FSU(5)$ superpotential for new inflation respecting the $U(1)_R \times Z_n$ symmetry can be written as,
\bea	
		W_{FSU(5)} &=&  S \left(-\mu^2 + \dfrac{(10_H\overline{10}_H)^m}{\Lambda^{2 m-2}}\right)\notag\\
		&+&\lambda_1 10_H 10_H 5_h + \lambda_2 \overline{10}_H\overline{10}_H\overline{5}_h \notag\\
		&+& y_{i j}^{(d)} 10_i 10_j 5_h + y_{i j}^{(u,\nu)} 10_i \overline{5}_j \overline{5}_h + y_{i j}^{(e)} 1_i \overline{5}_j 5_h,  \label{superpot2}
\eea
where $n=m$ for odd $n$ and $n=2m$ for even $n$ in accordance with the charge assignments of Table-I. The first-line terms of the above superpotential are relevant for new inflation which is realized by the scalar SM gauge-singlet components ($\Phi,\,\overline{\Phi}$) of ($10_H$, $\overline{10}_H$) pair. See \cite{Ellis:2014xda,Gonzalo:2016gey} where the scalar SM gauge-singlet component of matter superfield $10_i$, i.e., the sneutrino field, has been employed to realize inflation in $FSU(5)$. The color triplets of $5_h$ and $\overline{5}_h$ attain GUT scale masses via interaction terms in the second line of Eq.~(\ref{superpot2}). The electroweak Higgs doublets remain massless as the bilinear term $5_h \overline{5}_h$ is forbidden by the $R$-symmetry. Thus the doublet-triplet splitting problem is readily solved in $FSU(5)$ model due to the missing partner mechanism \cite{Antoniadis:1987dx} whereas in $SU(5)$ model a fine tuned solution is assumed.  The MSSM $\mu$ problem is managed by the Giudice-Masiero mechanism \cite{Giudice:1988yz}. Finally, the Dirac mass terms of all the fermions are generated by the Yukawa couplings $y_{ij}^{(u)}$, $y_{ij}^{(d,e)}$ and $y_{ij}^{(\nu)}$  of the terms in the third line of Eq.~(\ref{superpot2}). As the mass terms for the right handed neutrinos are not $R$-symmetric we need to go beyond the minimal framework considered above to realize light neutrino masses in $FSU(5)$ model as  is briefly discussed in section-5 along with reheating and leptogenesis. Compared to the $FSU(5)$  model, this problem is elegantly solved in $SU(5)$ model.

To derive the required form of the inflationary scalar potential we consider the following part of the above superpotential,
\be
		W_{FSU(5)} \supset  S \left(-\mu^2 + \frac{(\Phi\overline{\Phi})^m}{\Lambda^{2 m-2}}\right),
\ee
where the superfields, $\Phi$ and $\overline{\Phi}$, respectively represent the SM gauge-singlet components of $10_H$ and $\overline{10}_H$ Higgs superfields. The global SUSY vacuum is given by
\be	
 \langle S \rangle = 0,\quad	\langle \overline{\Phi}  \Phi \rangle  =  M^2,
\ee
where the gauge symmetry breaking scale, $M$, is defined as
		\begin{equation}
		M=(\mu \Lambda^{m-1})^{\frac{1}{m}}.
		\end{equation}
The	K\"ahler potential for $FSU(5)$ model can be expanded as,
\bea
K &=&  \bigg(1 + k_2 w_2 + \cdots \bigg) \bigg( \lvert S \rvert ^2+ \lvert \Phi \rvert ^2+ \lvert \overline{\Phi} \rvert^2 
+\frac{\kappa_S}{4}\frac{ \lvert S \rvert ^4}{m_P^2}+\frac{\kappa_{\Phi}}{4}\frac{ \lvert \Phi \rvert ^4}{m_P^2}+\frac{\kappa_{\bar{\Phi}}}{4}\frac{ \lvert \overline{\Phi} \rvert ^4}{m_P^2} \bigg. \notag \\
&+& \bigg. \kappa_{S\Phi}\frac{ \lvert S \rvert ^2 \lvert \Phi \rvert ^2}{m_P^2}+\kappa_{S\bar{\Phi}}\frac{ \lvert S \rvert ^2 \lvert \overline{\Phi} \rvert ^2}{m_P^2}+\kappa_{\Phi \bar{\Phi}}\frac{ \lvert \Phi \rvert ^2 \lvert \overline{\Phi} \rvert ^2}{m_P^2} + \cdots \bigg),
\eea
where,
\be
w_2 = \frac{\left(\Phi\overline{\Phi}\right)^m}{\Lambda^{2 m}} + \text{h.c.}.
\ee
Now using Eq.~(\ref{potformula}) with $z_i=\{S,\Phi,\overline{\Phi},...\}$, the scalar potential for $|S|,|\Phi|\ll m_P$ becomes in the D-flat direction ($\overline{\Phi} = \Phi^*$) as, 
\begin{equation}
V \simeq \mu^4 \left( 1 - \left| \frac{\Phi}{M} \right|^{2m}\right)^2 
+ \mu^{4} \bigg| \frac{2 m \Phi^{2 m-1}}{M^{2m}} \bigg|^{2} | S |^{2}
+\mu^4 \left(-\kappa_S\frac{ \lvert S \rvert ^2}{m_P^2}+2(1-\kappa_{10})\frac{|\Phi|^2}{m_P^2}\right),
\end{equation}	
where the coupling, $\kappa_{10} =(\kappa_{S\Phi}+\kappa_{S\bar{\Phi}})/2$. Note that the choice of D-flat condition automatically removes the complex phase dependence of $\Phi$ in this case. As discussed earlier, the $S$ field can be stabilized to zero by assuming $\kappa_S<-1/3$. Therefore, the $FSU(5)$ scalar potential takes the following form with the canonically normalized real field $\phi\equiv 2 \text{Re}|\Phi|$,
\begin{equation}
V_{FSU(5)} \simeq \mu^4\left( 1 - \frac{\gamma}{2} \frac{\phi^2}{m_P^2} - 2 \bigg(\frac{\phi}{2M}\bigg)^{2m} + \bigg(\frac{\phi}{2M}\bigg)^{4m} \right),  \label{vfsu5}
\end{equation}
where, $\gamma\equiv\kappa_{10}-1 \geq 0$. Except for the factor of $2$ instead of $\sqrt{2}$ with $M$ and $2m$ instead of $m$ the above potential is similar to the $SU(5)$ potential in Eq.~(\ref{vsu5}). Therefore, in the analytical discussion below we assume $\sqrt{2}M = 1$ units for $V_{SU(5)}$ and  $2M = 1$ units for $V_{FSU(5)}$ and exchange $m \leftrightarrow 2m$ in order to switch between the two models $SU(5) \leftrightarrow FSU(5)$.

\section{\large{\bf Results and discussion}}
In new inflation models, inflation occurs below the vacuum expectation value (VEV) where inflaton field is assumed to start from somewhere close to the origin ($\phi = 0$) and rolls down the minimum of the potential at $\phi=1$. This initial condition of inflaton can be realized dynamically with an earlier stage of preinflation  \cite{Senoguz:2004ky,Izawa:1997df,Yamaguchi:2004tn}. For example, this early stage of preinflation, characterized by an energy scale $\mu_{\text{pre}} \gg \mu$, can be realized by a model of supersymmetric hybrid preinflation where a gauge-singlet field $X$ plays the role of preinflaton. This stage provides a large effective mass $\sim \mu_{\text{pre}}^4/m_P^2$ to the inflaton, $\phi$, during preinflation. The $\phi$-field immediately  stabilizes to its minimum which becomes, $\phi_{\text{min}} \simeq (\mu/\mu_{\text{pre}})^2 X_c$, with $X_c \sim \mu_{\text{pre}}$ at the end of preinflation \cite{Senoguz:2004ky,Izawa:1997df,Yamaguchi:2004tn}. After the end of preinflation the $\phi$-field destabilizes to realize new inflation from its initial value $\phi_{\text{min}}$, now lying reasonably close to the origin. This way $\phi$ field acquires its natural initial conditions suitable for new inflation.

We now derive the slow-roll predictions for $SU(5)$ model of new inflation with $\sqrt{2}M=1$ units. In order to obtain the predictions of $FSU(5)$ model from the predictions of $SU(5)$ model we need to replace $m\rightarrow 2\,m$ and $M \rightarrow\sqrt{2}M$. Our obvious expectation is to obtain twice as many solutions in $SU(5)$ model as in $FSU(5)$ model or in \cite{Senoguz:2004ky}. The leading order slow roll parameters for the $SU(5)$ potential in Eq.~(\ref{vsu5}) are given as,
\bea
\epsilon (\phi) &=&\frac{m_P^2}{2} \left( \frac{\partial_{\phi} V_{SU(5)}}{V_{SU(5)}} \right)^2 \simeq \dfrac{1}{2}\bigg(\frac{\gamma\phi}{m_{P}}+mm_{P}\phi^{m-1}\bigg)^{2} ,\\
\eta (\phi) &=& m_P^2 \left( \frac{\partial_{\phi}^2 V_{SU(5)}}{V_{SU(5)}} \right)  \simeq -\gamma-m(m-1)m_{P}^{2}\phi^{m-2},
\eea
where we have defined, $\partial_{\phi} \equiv \frac{\partial}{\partial \phi}$ and $\partial_{\phi}^2 \equiv \frac{\partial^2}{\partial \phi^2}$. In general $|\eta(\phi)| \gg \epsilon (\phi)$ for $\phi \ll 1$. The end of inflation occurs when $ \eta(\phi_e) \simeq -1 $, where the field value at the end of inflation, $\phi_e$, is obtained as,
\begin{equation}
\phi_e\simeq\bigg[\dfrac{1-\gamma}{m(m-1)m_{P}^{2}}\bigg]^{\dfrac{1}{m-2}}. \label{phie}
\end{equation}
The $N_0$ number of $e$-folds before the end of inflation is given by
\bea
N_0 &=& \frac{2}{m_P^2}\int_{\phi_e}^{\phi_0} \left( \frac{V_{SU(5)}}{\partial_{\phi}V_{SU(5)}} \right) d\phi \notag \\
&\simeq& \left\{\begin{array}{c}\dfrac{-1}{(2m-2)\gamma}\log\left[\dfrac{\phi_0^{2m-2}}{\gamma+4 m m_P^2 \phi_0^{2m-2}}\dfrac{\gamma+4 m m_P^2 \phi_e^{2m-2}}{\phi_e^{2m-2}}\right],\qquad \text{$\gamma \geq 0$} \\\\
\dfrac{1}{4 m m_P^2 (2m-2)}\left(\frac{1}{\phi_0^{2m-2}}-\frac{1}{\phi_e^{2m-2}}\right), \qquad \qquad\qquad\qquad\qquad\text{$\gamma \rightarrow 0$} \\
\end{array}\right.
\eea
where $\phi_0$ is the field value at horizon exit of the comoving scale $l_0$ and before the last $N_0$ number of e-folds. Here, the expression of $N_0$ derived in the small $\gamma$ limit matches with the one obtained in \cite{Alabidi:2005qi}. Using Eq.~(\ref{phie}), we can solve above equations for $\phi_0$ as
	\begin{equation}
	\phi_0^{2m-2}\simeq\left\{\begin{array}{c}
	\left(\dfrac{2m-1}{1+(2m-2)\gamma}\right)\dfrac{\gamma(1-\gamma)}{4 m (2 m-1) m_P^2}e^{-(2m-2)\gamma N_0},\quad \,\,\,\, \text{$\gamma\gtrsim \gamma_0$} \\\\
	
	\dfrac{1}{4 m m_P^2 (2m-2)}\left(\dfrac{1}{N_0}\right), \qquad \qquad\qquad\qquad\qquad\qquad\text{$\gamma \ll \gamma_0$} \\
	\end{array}\right.    \label{phi0}
	\end{equation}
where to correctly describe the valid limits of above approximate results we have defined $\gamma_0$ as,
	\begin{equation}
	\gamma_0 \equiv \dfrac{\log(2m-1)}{2(m-1)N_0}.
	\end{equation}

\begin{figure}[t]
	\centering 	\includegraphics[width = 8.0cm]{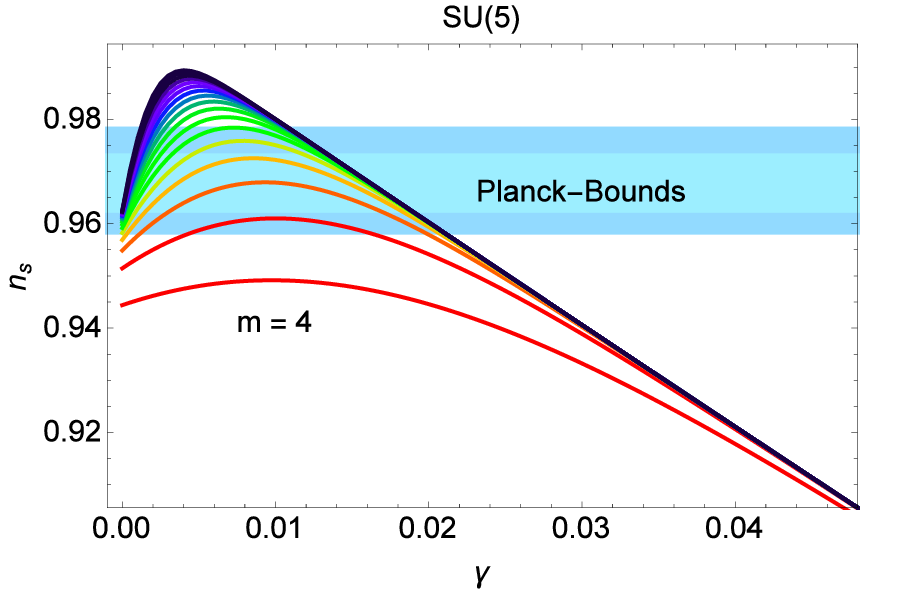} 
	\centering 	 \includegraphics[width = 8.0cm]{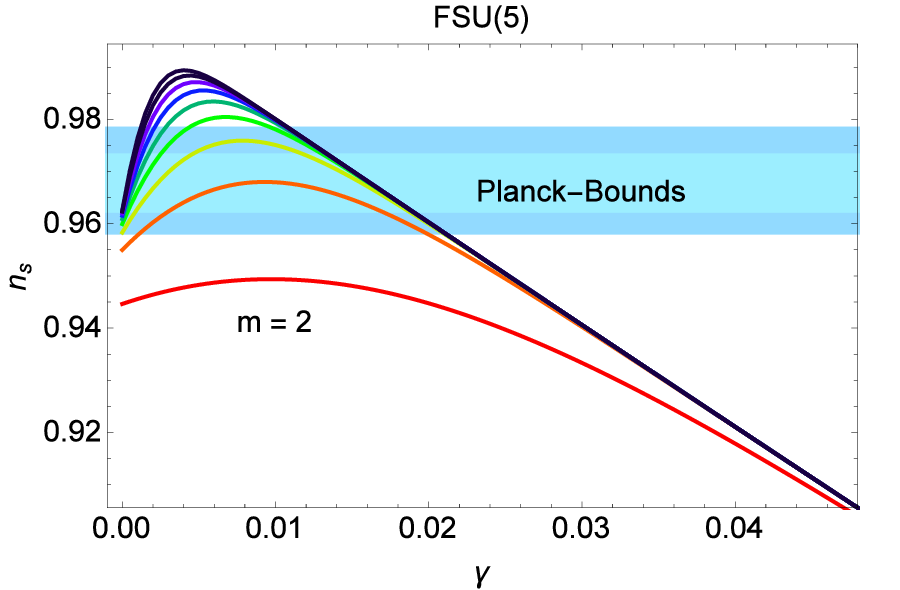}
	\caption{The scalar spectral index, $n_s$, versus the coupling, $\gamma$, for $SU(5)$ (left panel) and Flipped $SU(5)$ (right panel). The value of $m$ increases in integer steps from bottom to top and it starts from 4 in case of $SU(5)$ and 2 for Flipped $SU(5)$. The lighter and darker blue bands represent Planck's 1-$\sigma$ and 2-$\sigma$ bounds respectively.} \label{nsg}
\end{figure}
\begin{figure}[t]
\centering	\includegraphics[width=8.0cm]{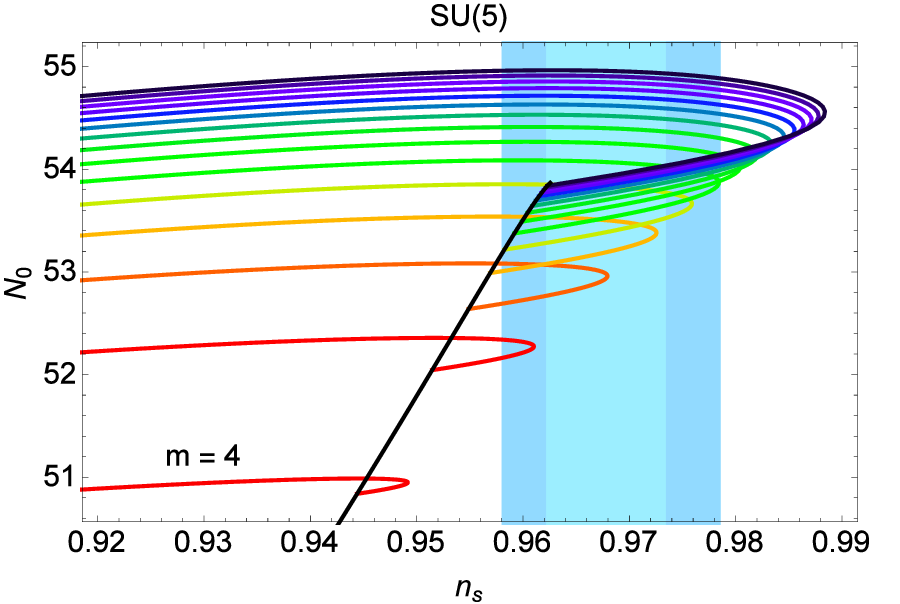}
\centering	\includegraphics[width=8.0cm]{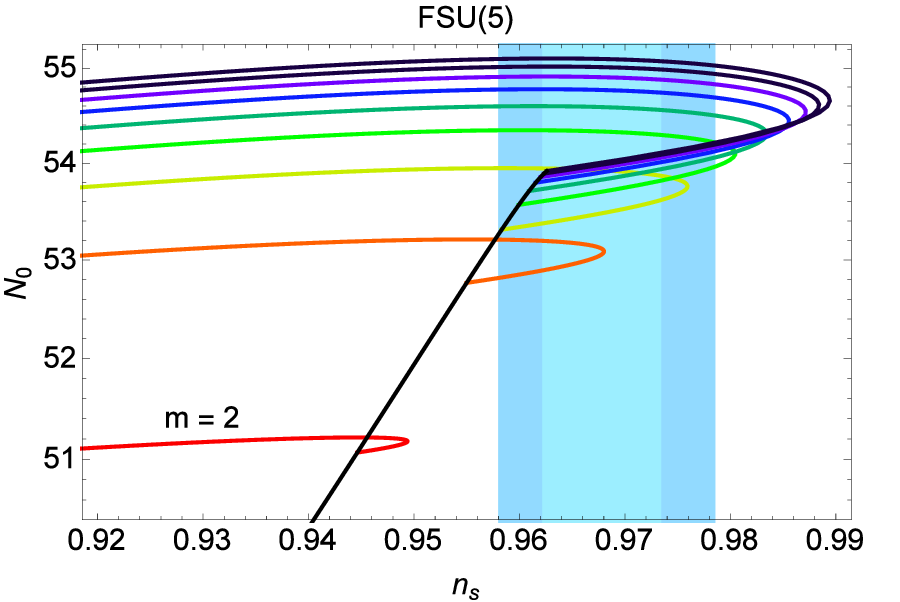} 
\centering	\includegraphics[width=8.0cm]{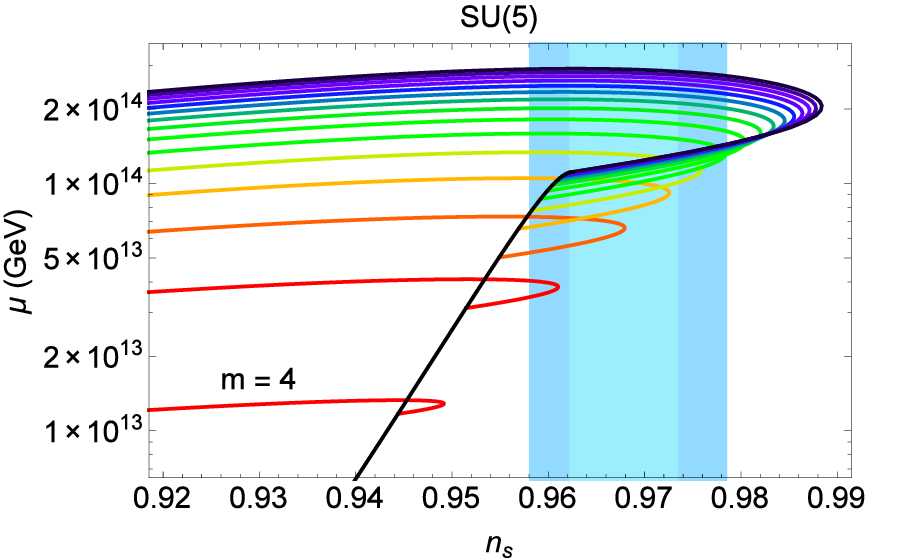} 
\centering 	\includegraphics[width=8.0cm]{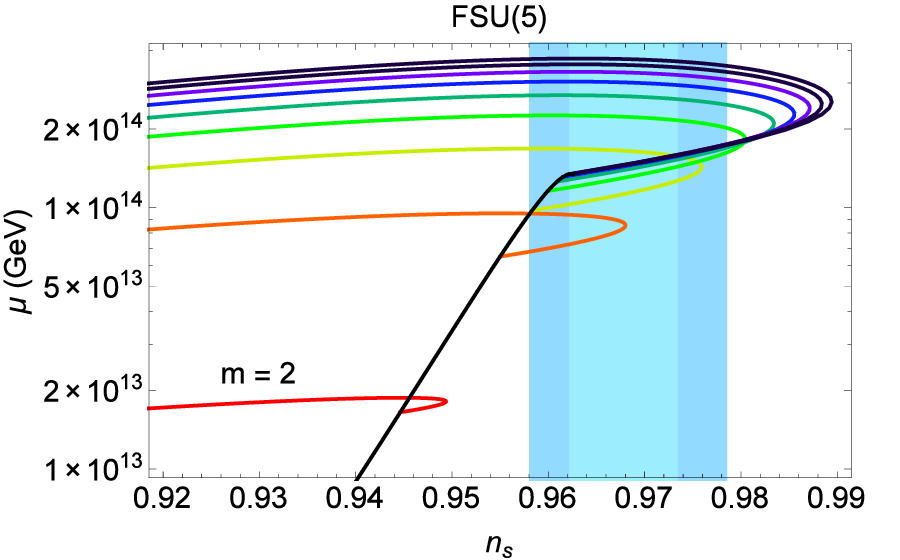} 
	\caption{The number of e-folds, $N_0$, and the energy scale of inflation, $\mu$, versus the scalar spectral index, $n_s$, for $SU(5)$ (left panel) and Flipped $SU(5)$ (right panel). The value of $m$ increases in integer steps from bottom to top and it starts from 4 (2) for $SU(5)$(Flipped $SU(5)$). The black line represents the case $\gamma=0$. The lighter and darker blue bands represent Planck's 1-$\sigma$ and 2-$\sigma$ bounds respectively.}%
	\label{nemuns}%
\end{figure}

The scalar spectral index, $n_s$, the first most important discriminator of inflationary models, can now be expressed in terms of $\gamma$ and $N_0$ as,
\begin{equation}
n_s \simeq 1 + 2 \eta (\phi_0) \simeq \left\{\begin{array}{c}1-2\gamma\left(1+\left(\dfrac{(2m-1)(1-\gamma)}{1+(2m-2)\gamma}\right)e^{-(2m-2)\gamma N_0}\right),\quad \text{$\gamma\gtrsim \gamma_0$} \\\\
1-\dfrac{(2m-1)}{(m-1)}\dfrac{1}{N_0}. \qquad \qquad\qquad\qquad\qquad\qquad\quad\qquad \text{$\gamma \ll \gamma_0$} \\
\end{array}\right.
\end{equation}
The same expression of $n_s$ in the small $\gamma$ limit is also obtained in \cite{Alabidi:2005qi}.
Note that in all our numerical calculations we have included next to leading order slow-roll expressions \cite{Stewart:1993bc}. The above approximate results of $n_s$ for $SU(5)$ model are consistent with our numerical estimates shown in the left panel of Fig.~(\ref{nsg}) for $4 \leq m \leq 20$ in integer steps. For $FSU(5)$ model, the numerical results for $n_s$ are shown in the right panel of Fig.~(\ref{nsg}) for $2 \leq m \leq 10$ in integer steps. The predictions of $n_s$ are shown to be compatible with the recent Planck's data bounds on $n_s$ for $0\lesssim \gamma \lesssim 0.02$ with $m \geq 5$ ($m \geq 3$) for $SU(5)$  ($FSU(5)$) model. However, for larger values of $m$, $n_s \gtrsim 0.98$ and lies outside the Planck's 2-$\sigma$ bound for $0.002\lesssim \gamma \lesssim 0.01$. For $\gamma \gtrsim 0.02$, the scalar spectral index $n_s \lesssim 0.96$ and lies outside the Planck's 2-$\sigma$ bound for all values of $m$.

In estimating the predictions of the various inflationary parameters we have implicitly assumed a standard thermal history of the Universe after inflation with the following expression for $N_0$ \cite{Kolb:1990vq},
\begin{equation}
N_{0}\simeq54+(1/3)\ln(T_{r}/10^{9}\text{ GeV})+(2/3)\ln(\mu/10^{14} \text{ GeV}),
\end{equation}
where $T_r$ is the reheat temperature. The numerical estimates of reheat temperature with a realistic non-thermal leptogenesis is discussed in the next section.
The number of $e$-folds $N_0$ consistent with Planck bounds on $n_s$ turns out to be $51-52$ as shown in the Fig.~(\ref{nemuns}).
Above expression of $N_0$ also depends on $\mu$, the energy scale of inflation, which can be calculated from the expression of the amplitude of scalar perturbation,
\begin{equation}
A_s(k_0)=\frac{1}{24 \pi^2} \left( \dfrac{V_{SU(5)}/m_P^4}{\epsilon} \right)_{\phi=\phi_0} 
\simeq\frac{1}{12\pi^2 m_P^6}\left(\dfrac{\mu^2}{4 m \phi_0^{2m-1}+\gamma \frac{\phi_0}{m_P^{2}}}\right)^2, 
\end{equation}
where  $A_s(k_0) = 2.196 \times 10^{-9}$ at the pivot scale $k_0 = 0.05$ Mpc$^{-1}$. Therefore, the energy scale of inflation is obtained as
\begin{equation}
\mu=\left({12\pi^2 m_P^6 A_s \left(4 m \phi_0^{2m-1}+\gamma \frac{\phi_0}{m_P^{2}}\right)^2}\right)^\frac{1}{4},
\end{equation}
\begin{figure}[t]
	\centering
	\includegraphics[width=8.0cm]{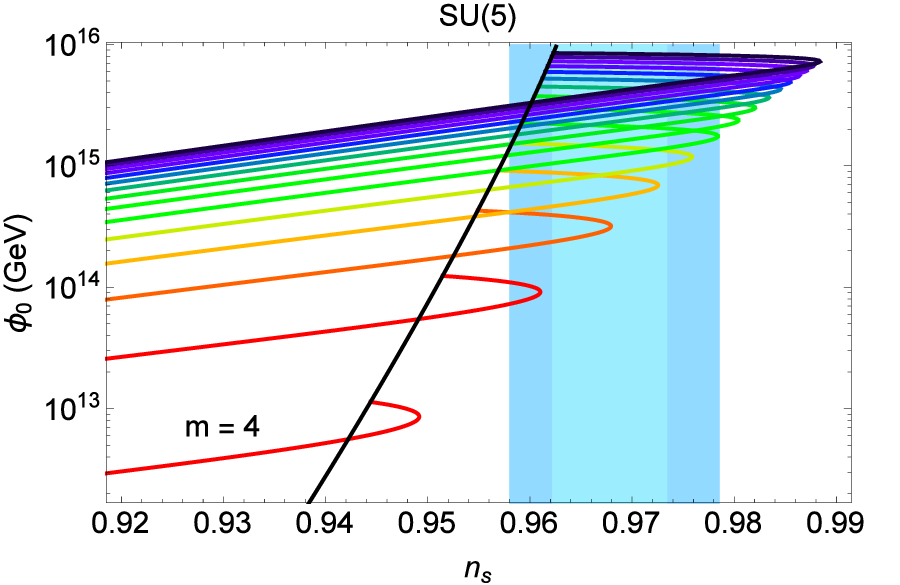}
    \includegraphics[width=8.0cm]{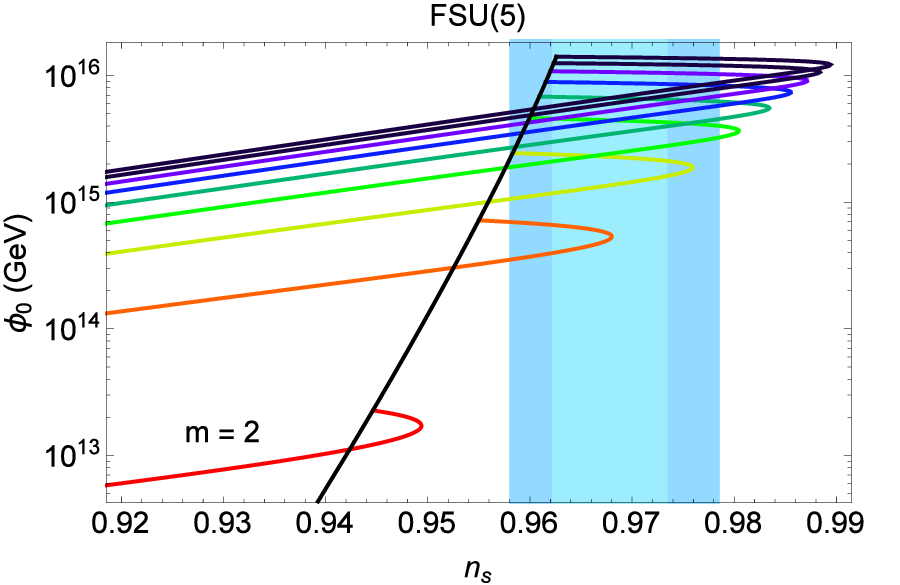}
	\includegraphics[width=8.0cm]{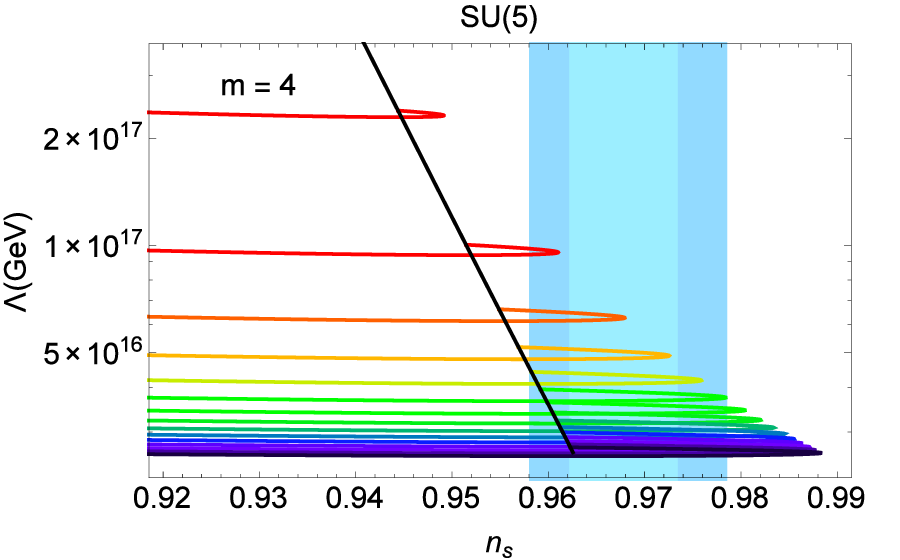} %
	\includegraphics[width=8.0cm]{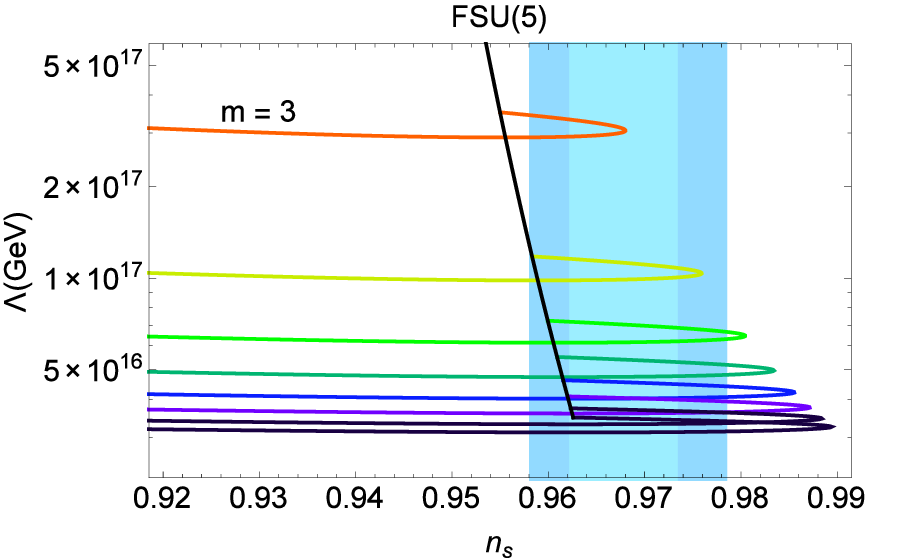} %
	\caption{The field value at the pivot scale, $\phi_0$, and the cutoff scale, $\Lambda$, versus the scalar spectral index, $n_s$, for $SU(5)$ (left panel) and Flipped $SU(5)$ (right panel). The value of $m$ increases in integer steps from bottom to top and it starts from 4 (2) for $SU(5)$(Flipped $SU(5)$). The black line represents the case $\gamma=0$. The lighter and darker blue bands represent Planck's 1-$\sigma$ and 2-$\sigma$ bounds respectively.}%
	\label{phiLns}
\end{figure}
where $\phi_0$ as a function of $N_0$ is given in Eq.~(\ref{phi0}). The numerical predictions for the energy scale of inflation, $5.5 \times 10^{13} \text{ GeV } \lesssim \mu \lesssim 3 \times 10^{14}$ GeV, are obtained   within the Planck data bounds as shown in the lower panel of Fig.~(\ref{nemuns}) for the various values of $m$. These small values of $\mu \ll 2 \times 10^{16}$ GeV is a common feature of small field model of inflation where $\phi_0 \ll m_P$ \cite{Martin:2013tda}. 

In the upper panel of Fig.~(\ref{phiLns}) we display the field values at the pivot scale for which the range, $4 \times 10^{14} \text{ GeV } \lesssim \phi_0 \lesssim 10^{16}$ GeV, lies within the Planck data bounds on $n_s$. These values are not only far less than the Planck mass scale, $m_P=2\times10^{18}$ GeV, they also lie below the corresponding values of the cutoff scale, $2 \times 10^{16} \text{ GeV } \lesssim \Lambda \lesssim 2.5 \times 10^{17}$~GeV, as shown in the lower panel of Fig.~(\ref{phiLns}) for the various values of $m$. For very large values of $m$, the cutoff scale $\Lambda$ approaches the gauge symmetry breaking scale $M$ from above and the field value during inflation reaches $\sqrt{2} M$ ($2M$) from below for $SU(5)$ ($FSU(5)$) model. Therefore, for the values of $m$ considered in our numerical calculations, i.e., $4 \lesssim m \lesssim 20$ ($3 \lesssim m \lesssim 10$) in case of $SU(5)$ ($FSU(5)$), the ratio $\phi/\Lambda \lesssim 0.1$ is consistent with the central value of $n_s \simeq 0.968$. However, the large values of $m$ considered here are merely to show the trend of the various predictions in the large $m$ limit otherwise only first few values of $m$ correspond to a more realistic situation with the cutoff scale lying reasonably below $M$.

\begin{figure}[t]
	\centering
	\includegraphics[width=8.0cm]{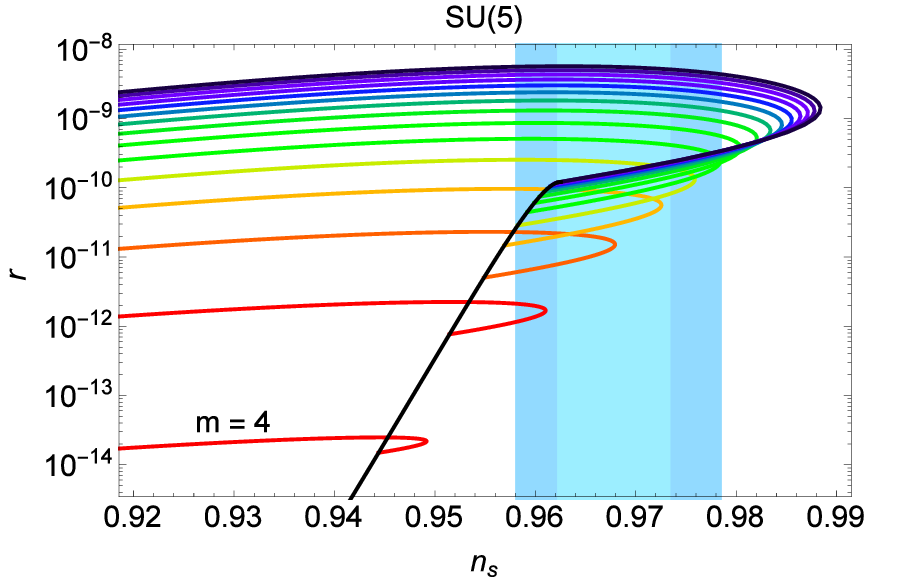}
	\includegraphics[width=8.0cm]{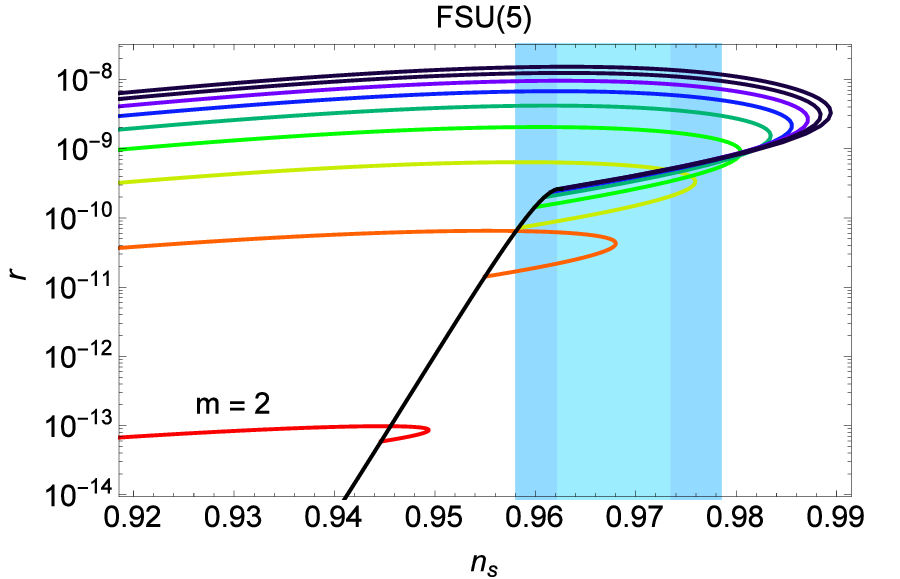}
	\includegraphics[width=8.0cm]{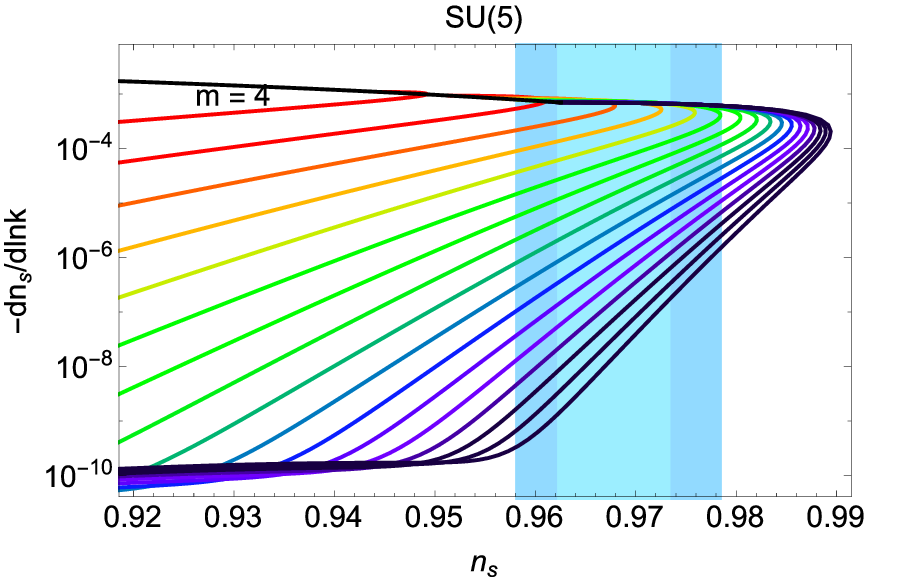}
	\includegraphics[width=8.0cm]{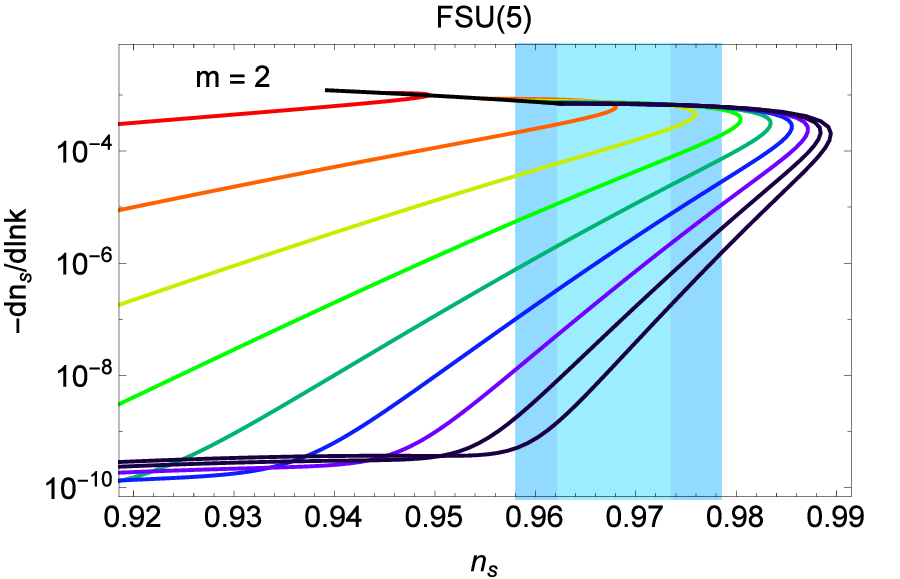} 
	\caption{The tensor to scalar ratio, $r$, and the running of spectral index, $dn_s/dlnk$, versus the scalar spectral index, $n_s$, for $SU(5)$ (left panel) and Flipped $SU(5)$ (right panel). The value of $m$ increases in integer steps from bottom to top and it starts from 4 (2) for $SU(5)$(Flipped $SU(5)$). The black line represents the case $\gamma=0$. The lighter and darker blue bands represent Planck's 1-$\sigma$ and 2-$\sigma$ bounds respectively.}%
	\label{rnsrunsns}
\end{figure}

The next important discriminator for the inflationary potentials is the tensor to scalar ratio, $r \simeq 16\,\epsilon(\phi_0)$, which is constrained by the recent Planck data to be, $r \lesssim 0.01$ \cite{Ade:2015lrj}. As depicted in the upper panel of Fig.~(\ref{rnsrunsns}), the range of the tensor to scalar ratio for both models turns out to be, $10^{-12} \lesssim r \lesssim 10^{-8}$, for $n_s$ lying within the current Planck data bounds. 
Again this is a generic feature of hilltop like small field models \cite{Boubekeur:2005zm}. The detection of primordial gravity waves in future experiments \cite{{Andre:2013afa,Matsumura:2013aja,Kogut:2011xw,Finelli:2016cyd}} with $r \gtrsim 10^{-4}$ can rule out these models. Finally, we consider the running of spectral index $dn_s/d\ln k$, which appears at the next to leading order slow-roll approximation (with $\epsilon \ll |\eta|$) as,
\be
\frac{dn_s}{d\ln k} \simeq  16 \epsilon(\phi_0) \eta(\phi_0) - 2\,\xi^2(\phi_0) \simeq -\frac{r (1 - n_s)}{2} - \sqrt{\frac{r}{2}} \, m (m-1)(m-2)\phi_0^{m-3} < 0, 
\ee
where, $\xi^2(\phi) = m_P^4 \left( \frac{\partial_{\phi} V \partial^3_{\phi} V}{V^2}\right)$. As shown in the lower panel of Fig.~(\ref{rnsrunsns}), we obtain the range, $7 \times 10^{-10} \lesssim |dn_s/d\ln k| \lesssim 10^{-3}$, in accordance with Planck data bounds on $n_s$. The relatively small values of the tensor to scalar ratio and the running of the scalar spectral index are consistent with the underlying assumptions of Planck data bounds on $n_s$ derived from the base $\Lambda$CDM model \cite{Ade:2015lrj}. This also justifies our comparison of the model predictions with the right choice of experimental data.

\section{Reheating with non-thermal leptogenesis}
In this section we mainly focus on the realization of reheating with leptogenesis. A successful baryogenesis is usually generated through the sphaleron processes \cite{Kuzmin:1985mm,Fukugita:1986hr,Khlebnikov:1988sr} where an initial lepton asymmetry, $n_L/s$, is partially converted into a baryon asymmetry, $n_B/s \simeq - 0.35\, n_L/s$ \cite{Khlebnikov:1988sr,Harvey:1990qw}. For reheating with leptogenesis, this requires the presence of lepton-number violating  interactions with inflaton. A typical such term in $SU(5)$ model can be expressed as
 \be
 W \supset \beta_{ij} 
 \bigg(\frac{(\text{Tr}\,\Phi^{\alpha_1})^{k_1}(\text{Tr}\,\Phi^{\alpha_2})^{k_2}...(\text{Tr}\,\Phi^{\alpha_n})^{k_n}}{\Lambda^{m-1}}\bigg) N_i^c N_j^c \supset  m\beta_{ij}\left( \frac{v}{\Lambda} \right)^{m-1} \delta \Phi N_i^c N_j^c ,  \label{trsu5}
 \ee
where $\sum_{i=1}^{n}k_i\alpha_i = m$ for the non-negative integer values for $\alpha_i$ and $k_i$. The leading order Yukawa interaction term is obtained by expanding the GUT Higgs field $\Phi$ about its VEV, $v$, with the perturbed field, $\delta \Phi$, defined as $(\text{Tr}\,\Phi^{\alpha_i})^{k_i} = v^{k_i \alpha_i} + k_i \alpha_i \,\delta \Phi$. The effective mass term for right-handed neutrinos is $m_{\nu_{ij}} + \beta_{ij}(v/\Lambda)^{m-1}v$ which after diagonalizing gives masses $M_1, M_2$ and $M_3$.

Another important term relevant for reheating is given as 
 \be
 W \supset \beta \bigg(\frac{(\text{Tr}\,\Phi^{\alpha_1})^{k_1}(\text{Tr}\,\Phi^{\alpha_2})^{k_2}...(\text{Tr}\,\Phi^{\alpha_n})^{k_n}}{\Lambda^{m-1}}\bigg) \bar{5}_h (\Phi^{\alpha_0})^{k_0} 5_h \supset  m \beta \left( \frac{v}{\Lambda} \right)^{m-1} \delta \Phi H_u  H_d,
\ee
with $\sum_{i=0}^{n}k_i\alpha_i = m$ while assuming non-negative integer values for $\alpha_i$ and $k_i$. This interaction term with electroweak doublet Higgs pair $(H_u\,H_d)$, as already included in the superpotential (in Eq.~\ref{superpot1}), can lead to inflaton decay into Higgsinos. A similar reheating scenario with gauge singlet inflaton has recently been employed in \cite{Rehman:2017gkm} for $\mu$-hybrid inflation. For the sake of simplicity, the effect of this interaction is neglected in our numerical analysis by assuming $\beta \ll \beta_{ij}$. 

In case of $FSU(5)$, the realization of successful reheating and leptogenesis can proceed with the following gauge invariant interaction term,
\be W \supset \beta_{ij} \left( \frac{10_{H} \overline{10}_H}{\Lambda^2}\right)^{m-2} \left(\frac{\overline{10}_H \, \overline{10}_H 10_i 10_j}{\Lambda} \right) \supset \beta_{ij} \left( \frac{\Phi \overline{\Phi}}{\Lambda^2}\right)^{m-2} 
\left(\frac{\overline{\Phi} \, \overline{\Phi}}{\Lambda}\right) N_i^c N_j^c, \label{eqn38}
\ee 
where $m=n$ for odd $n$ and $m=n/2$ for even $n$ in accordance with the charge assignments of Table-I. This term, however, is not $R$-symmetric according to the $R$-charge assignments of Table-I. To ameliorate the situation we can either assume the breaking of $R$-symmetry by higher order Planck scale suppressed operators (with $\Lambda=m_P$ in above equation), as suggested in \cite{Civiletti:2013cra}, or we can invoke a different $R$-charge assignment,
\be
q_R  (S,10_H,\overline{10}_H,5_h,\overline{5}_h,10_i,\overline{5}_i,1_i) = (1,0,0,1,1,1/2,-1/2,1/2),
\ee
which can allow this term in the above superpotential by making it R-symmetric. The price one must pay with this option is the disappearance of down-type quark Yukawa term in the superpotential, Eq.~(16), at tree-level. But the mass for down-type quark can be generated radiatively. The term in Eq.~(\ref{eqn38}) also generates an effective mass matrix for the right-handed neutrinos, $\beta_{ij}(M/\Lambda)^{2m-3}M$, or after diagonalization, $M_i$, and help us explain the tiny masses of neutrinos via see-saw mechanism.

At the end of inflation, the GUT Higgs fields ($\Phi$ in case of $SU(5)$ and $\Phi$, $\overline{\Phi}$ in case of $FSU(5)$) perform damped oscillations about SUSY minimum. The mass of oscillating inflaton is denoted by $m_{inf}$ and is given by,
\begin{equation}
m_{inf}= \left\{\begin{array}{c} \frac{m\,\mu^2}{M},\qquad \text{for } SU(5), \\\\
\frac{\sqrt{2}\,m\,\mu^2}{M}, \qquad \text{for } FSU(5). \\
\end{array}\right.
\end{equation}
\begin{figure}[ht]
	\centering 	\includegraphics[width = 8.0cm]{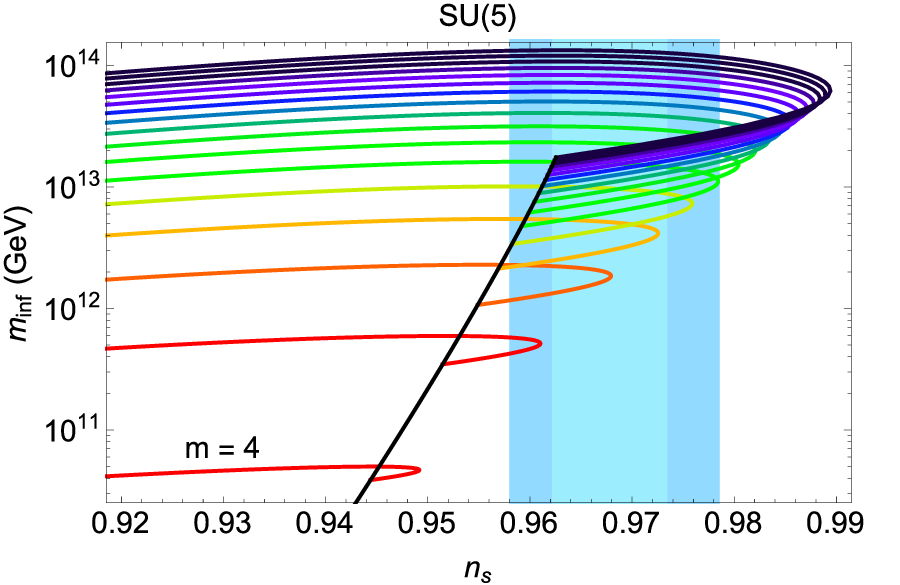} 
	\centering 	 \includegraphics[width = 8.0cm]{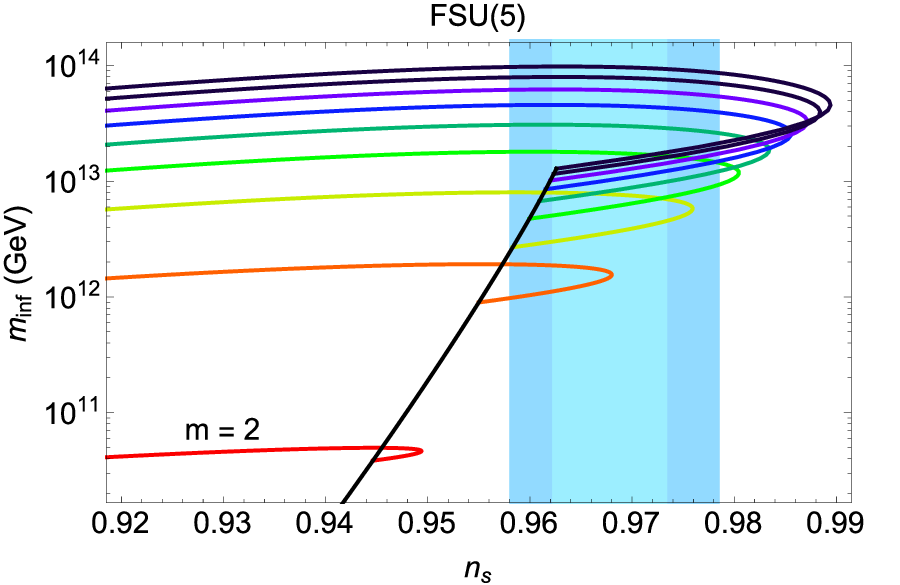}
	\caption{The inflaton mass, $m_{inf}$, versus the scalar spectral index, $n_s$, for $SU(5)$ (left panel) and Flipped $SU(5)$ (right panel). The value of $m$ increases in integer steps from bottom to top and it starts from 4 in case of $SU(5)$ and 2 for Flipped $SU(5)$. The lighter and darker blue bands represent Planck's 1-$\sigma$ and 2-$\sigma$ bounds respectively.} \label{minfns}
\end{figure}
The numerical estimate of inflaton mass for the various values of $m$ is shown in Fig.~(\ref{minfns}) for both $SU(5)$ and $FSU(5)$. The predicted range of inflaton mass, $ 3 \times 10^{11} \text{ GeV} \lesssim m_{inf} \lesssim 10^{14}$ GeV, lies within Planck bounds on $n_s$.  

The decay width of inflaton' decay through the effective Yukawa interactions described in Eq.~(\ref{trsu5}) and Eq.~(\ref{eqn38}) is given by
\begin{equation}
\Gamma \simeq \frac{1+y^2}{8\pi} \bigg(\frac{M_1}{M}\bigg)^2 m_{inf},
\end{equation}
where $y=M_1/M_2 \ll 1$ assuming a hierarchical pattern ($M_1 \leq m_{inf}/2 \ll M_2,\,M_3$) for the masses of right handed neutrinos. 
The reheat temperature $T_r$ is defined in terms of $\Gamma$ as 
\begin{equation}
T_r = \bigg(\frac{45}{2\pi^2g_*}\bigg)^{1/4}\sqrt{m_p \Gamma}, 
\end{equation}
where $g_* = 228.75$ for MSSM. 

In the limit $T_r <M_1 \leq m_{inf}/2 \ll M_{2,3}$ the lepton asymmetry, $n_{L}/s$,  is given by
\begin{equation}
\frac{n_L}{s} \simeq \frac{3}{2} \frac{T_r}{m_{inf}}\epsilon_1,
\end{equation}
where the CP asymmetry factor, $\epsilon_1$, is generated from the out of equilibrium decay of the lightest right-handed neutrino $N_1$ and is given by \cite{Luty:1992un,Covi:1996wh}
\begin{equation}
\epsilon_1 = -\frac{3}{8\pi} \frac{1}{(y_{(\nu)} \,y^{\dagger}_{(\nu)})_{11}}\sum_{i=2,3} \text{Im} \left[ \left( y_{(\nu)} \, y^{\dagger}_{(\nu)})_{1i} \right)^2 \right]\frac{M_1}{M_i}.
\end{equation}
Here the coupling $y_{\nu}$ represents the neutrino Yukawa coupling defined in Eqs.~(\ref{superpot1}) and (\ref{superpot2}) for $SU(5)$ and $FSU(5)$ respectively. Further assuming a normal hierarchical pattern of light neutrino masses the CP asymmetry factor, $\epsilon_1$, becomes \cite{Hamaguchi:2002vc}
\begin{equation}
\epsilon_1 = -\frac{3}{8\pi}\frac{m_{\nu_3} M_1}{v_{u}^2}\delta_{eff}, 
\end{equation}
where $m_{\nu_3}$ is the mass of the heaviest light neutrino, $v_{u}=\langle H_u \rangle $ is the vev of the up-type electroweak Higgs and $\delta_{eff}$ is the CP-violating phase. In the numerical estimates discussed below we take $m_{\nu_3} = 0.05$ eV, $|\delta_{eff}|=1$ and $v_u = 174$ GeV assuming large $\tan \beta $ limit.

The non-thermal production of lepton asymmetry, $n_{L}/s$, is given by the following expression  
\begin{equation}
\frac{n_L}{s} \lesssim 3 \times 10^{-10} \frac{T_r}{m_{inf}}\left(\frac{M_1}{10^6 \text{ GeV}}\right),
\end{equation}
with $M_1 \gg T_r $. Using the experimental value of $n_L/s\approx 2.5\times 10^{-10}$ and eliminating $M_1$ from above equations we obtain the following lower bound on $T_r$,
\begin{equation}
T_r \gtrsim 1.6\times 10^7 \text{GeV} \left(\frac{10^{16}\text{GeV}}{M}\right)^{1/2}\left(\frac{m_{inf}}{10^{11}\text{GeV}}\right)^{3/4}.
\end{equation}
\begin{figure}[h]
	\centering 	\includegraphics[width = 8.0cm]{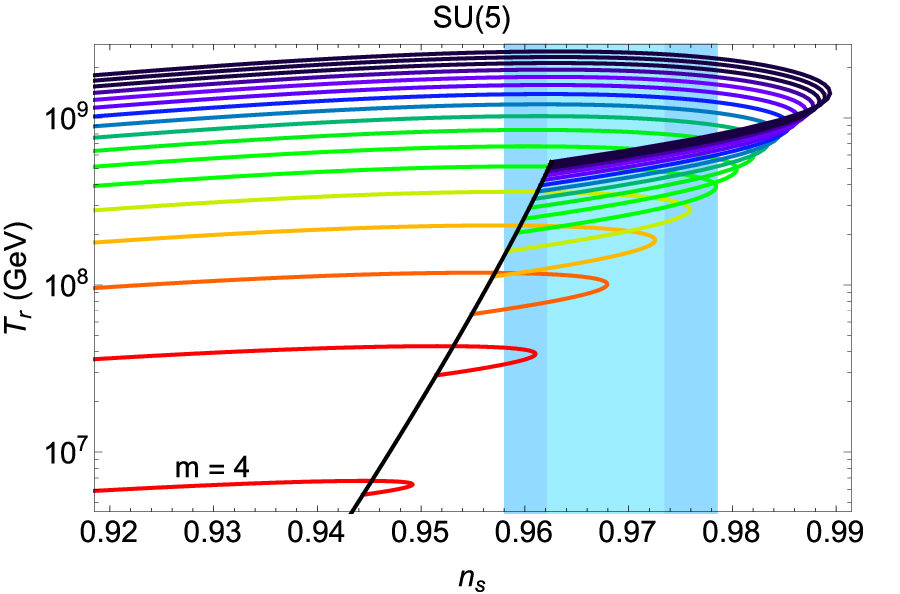} 
	\centering 	 \includegraphics[width = 8.0cm]{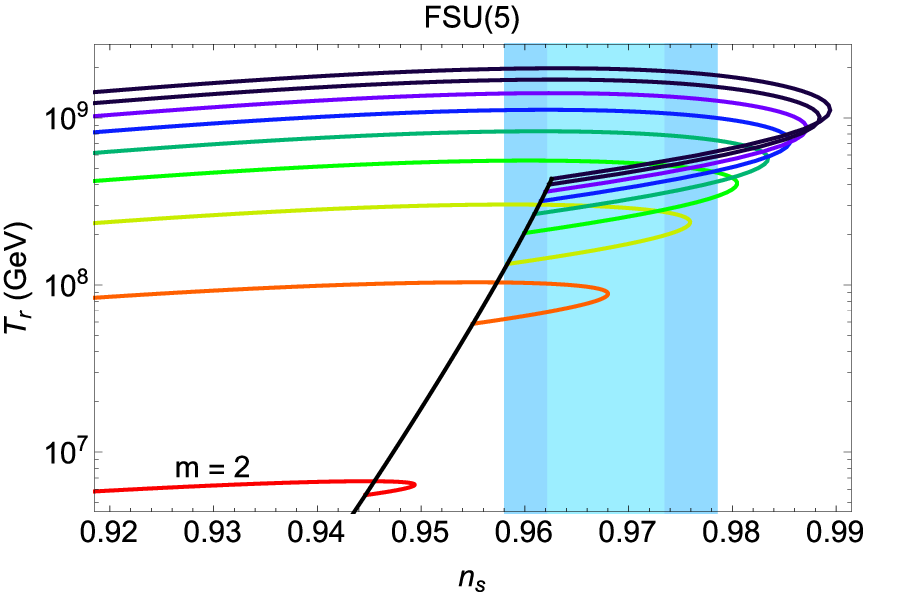}
	\caption{The reheat temperature, $T_r$, versus the scalar spectral index, $n_s$, for $SU(5)$ (left panel) and Flipped $SU(5)$ (right panel). The value of $m$ increases in integer steps from bottom to top and it starts from 4 in case of $SU(5)$ and 2 for Flipped $SU(5)$. The lighter and darker blue bands represent Planck's 1-$\sigma$ and 2-$\sigma$ bounds respectively.} \label{Tns}
\end{figure}

The variation of reheat temperature, $T_r$, versus  the scalar spectral index, $n_s$, is displayed in Fig.~(\ref{Tns}) for the various values of $m$. The  reheat temperature saturates around $2 \times 10^9$ GeV in the large $m$ limit. The range of reheat temperature, $3 \times 10^{7}\text{ GeV }\lesssim T_r \lesssim 2 \times 10^{9}$  GeV, is found consistent with the Planck bounds on $n_s$ and also avoids the gravitino problem \cite{Khlopov:1984pf} for $m_{3/2} \gtrsim 10$ TeV \cite{Kawasaki:2004qu,Kawasaki:2008qe,Kawasaki:2017bqm}.

\begin{figure}[H]
	\centering 	\includegraphics[width = 8.0cm]{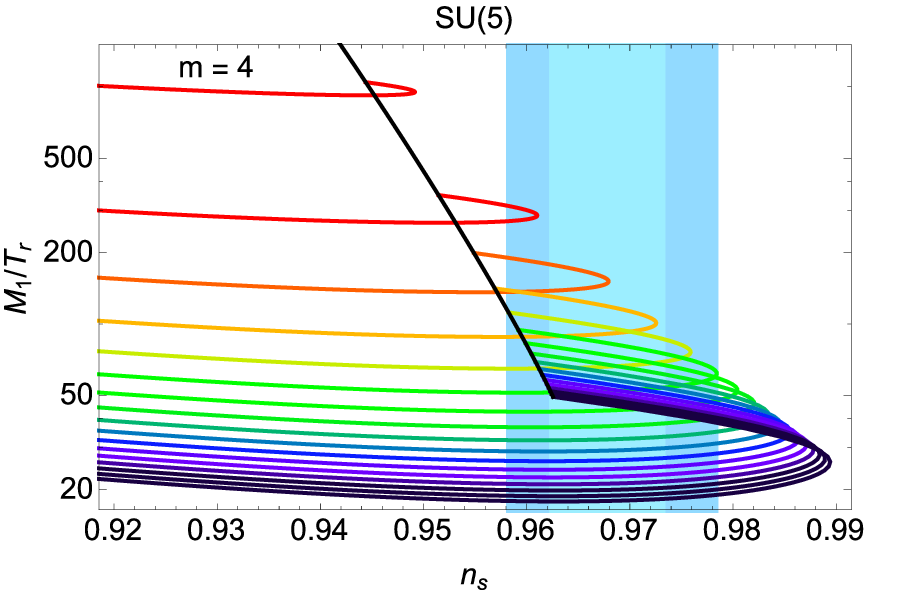} 
	\centering 	 \includegraphics[width = 8.0cm]{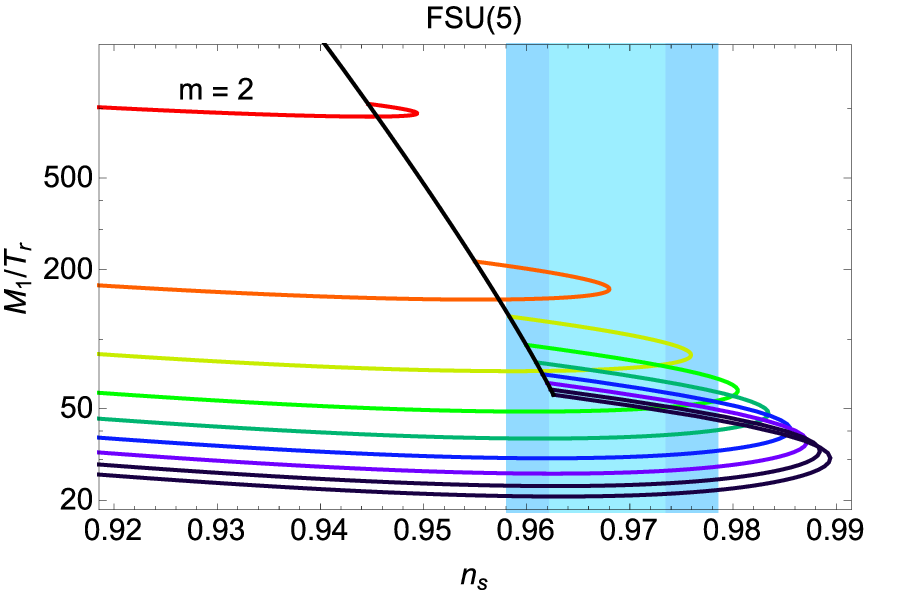}
	\caption{The ratio of the lightest right-handed neutrino mass to the reheat temperature, $M_1/T_r$, versus the scalar spectral index, $n_s$, for $SU(5)$ (left panel) and Flipped $SU(5)$ (right panel). The value of $m$ increases in integer steps from top to bottom. The lighter and darker blue bands represent Planck's 1-$\sigma$ and 2-$\sigma$ bounds respectively.} \label{mnuTns}
\end{figure}

Finally, in order to suppress the lepton asymmetry washout factor (which is proportional to $e^{-M_1/T_r}$), the lightest right handed neutrino $M_1$ is taken to be heavier than the reheat temperature $T_r$ ($M_1 \gg T_r$). Therefore, the washout effects can be safely ignored if $M_1/T_r \gtrsim 10$. As demonstrated in Fig. (\ref{mnuTns}), the ratio, $M_1/T_r$, is always greater than 10 for all values of $m$ which are consistent with Planck bounds on $n_s$. Thus, the observed lepton asymmetry can be generated for both $SU(5)$ and $FSU(5)$.

\section{Proton decay in $R$-symmetric SU(5) and FSU(5)}
Let us first discuss the status of proton decay in $R$-symmetric $SU(5)$ model. Here, the dimension four proton decay operators arising from the interactions at renormalizable level, $10_i \overline{5}_j \overline{5}_k \supset Q_i D^c_j L_k + U^c_i D^c_j D^c_k + E^c_i L_j L_k $, are not allowed by the $R$-charge assignments defined in Eq.~(\ref{Rsu5}). The same $R$-symmetry, however, allows the dimension five proton decay operators. For example, the renormalizable interaction terms in Eq.~(\ref{superpot1}), $y_{ij}^{(u)}10_{i}10_{j}5_h + y_{ij}^{(d,e)}10_{i}\overline{5}_{j}\overline{5}_h + \mu_3 H_C \overline{H}_C$, are relevant for the fast proton decay. Here, $\mu_3$ is a GUT scale mass parameter of the color-triplet Higgs pair ($H_C,\, \overline{H}_C$). These terms can potentially generate the dimension five fast proton decay (via $p \rightarrow K^+ \overline{\nu}$) in contradiction with the latest Super-Kamiokande data \cite{Takhistov:2016eqm}. For a recent update on the status of proton decay in the minimal SUSY $SU(5)$ GUT, see \cite{Ellis:2019fwf}. To ameliorate this problem of fast proton decay we assume somewhat higher mass values for squarks and sleptons of order $10$ TeV or so \cite{Hisano:2013exa} while respecting the current bounds on sparticle masses by LHC searches \cite{LHC}. For another possible dimension five proton decay we consider the operators at the non-renormalizable level, $10_i 10_j 10_k \overline{5}_l \supset Q_i Q_j Q_k L_l + U^c_i U^c_i D^c_j E^c_k$. With a suppression factor of order, $1/\Lambda$, the decay rate induced by these operators is adequately suppressed with $\Lambda \gtrsim 2 \times 10^{16}$ GeV and sparticle masses of order 10 TeV. Lastly, the dimension six proton decay operators do not pose any threat as their predictions lie well beyond the scope of Hyper-Kamiokande \cite{HKTDR}.

We now briefly discuss the status of proton decay in $R$-symmetric $FSU(5)$ model. For a similar discussion of proton decay in the context of SUSY hybrid inflation, see \cite{Kyae:2005nv,Rehman:2009yj,Rehman:2018nsn}. Regarding dimension four proton decay operators, it is important to note that the $SU(5)$ gauge invariant fast proton decay operator, $10_i \overline{5}_j \overline{5}_k$, is not gauge invariant under $FSU(5)$. The Yukawa terms, $y_{i j}^{(d)} 10_i 10_j 5_h + y_{i j}^{(u,\nu)} 10_i \overline{5}_j \overline{5}_h + y_{i j}^{(e)} 1_i \overline{5}_j 5_h$, can lead to the dimension five fast proton decay if supplemented by the $\left( \frac{(10_H\overline{10}_H)^2}{\Lambda^3}\right) 5_h\overline{5}_h$ term. Even though this term is $FSU(5)\times Z_n$ invariant, it is not $R$-symmetric. 

The dangerous dimension five proton decay operators at non-renormalizable level are contained in 
\bea
\left( \frac{10_H\overline{10}_H}{\Lambda^{2}} \right)^{m-2} \frac{10_i 10_j 10_k \overline{5}_l}{\Lambda} 
&\supset& \left( \frac{M}{\Lambda} \right)^{2m-4} \frac{Q_i Q_j Q_k L_l}{\Lambda}, \\ 
\frac{10_i \overline{5}_j \overline{5}_k \overline{1}_l }{\Lambda^2} \left( \frac{10_H\overline{10}_H}{\Lambda} \right)^{m-2} &\supset& \left( \frac{M}{\Lambda} \right)^{2m-4} \frac{D^c_i U^c_j U^c_k E^c_l}{\Lambda},
\eea
which are not invariant under the $R$-symmetry defined in Table-I. But, as mentioned in the previous section, if we either assume the breaking of $R$-symmetry by Planck suppressed operators or consider the $R$-symmetry defined in Eq.~(\ref{eqn38}), these operators with coefficient of order one can predict fast proton decay, for example, for $m=2$ or $n=4$. 
Now consider the following $U(1)_R \times Z_n$-symmetric operators:
\be
\frac{S 10_i 10_j 10_k \overline{5}_l}{\Lambda^2} \left( \frac{10_H\overline{10}_H}{\Lambda^{2}} \right)^{m-2} \quad \text{ and } \quad 
\frac{S 10_i \overline{5}_j \overline{5}_k \overline{1}_l }{\Lambda^2} \left( \frac{10_H\overline{10}_H}{\Lambda^2} \right)^{m-2}.
\ee
These operators are heavily suppressed due to $\langle S \rangle/\Lambda$ factor as $S$ field acquires a non-zero VEV, $\langle S \rangle \sim m_{3/2}(M/\mu)^2$, proportional to the gravitino mass, $m_{3/2} \gtrsim 10$ TeV, owing to softly broken SUSY \cite{Dvali:1997uq,King:1997ia}. Similarly, the renormalizable operators, $Q_i D^c_j L_k$, $D^c_i D^c_j U^c_k$ and $L_i L_j E^c_k$, may appear in the following $U(1)_R \times Z_n$-symmetric non-renormalizable operators:
\bea
\frac{S 10_H 10_i 10_j \overline{5}_k}{\Lambda^2} \left( \frac{10_H\overline{10}_H}{\Lambda^{2}} \right)^{m-2} &\supset& \frac{\langle S \rangle}{\Lambda}
\left(\frac{M}{\Lambda} \right)^{2m-3}  \left( Q_i D^c_j L_k + D^c_i D^c_j U^c_k \right), \\ 
\frac{S 10_H \overline{5}_i \overline{5}_j \overline{1}_k }{\Lambda^2} \left( \frac{10_H\overline{10}_H}{\Lambda^{2}} \right)^{m-2}  &\supset& \frac{\langle S \rangle}{\Lambda}
\left(\frac{M}{\Lambda} \right)^{2m-3} \left( L_i L_j E^c_k \right).
\eea
However, these operator are not allowed by the $Z_2$ matter parity described in Table-I. Thus, the proton decay through dimension five operators is adequately suppressed in $R$-symmetric $FSU(5)$ model and  can proceed through the dimension six operators mediated by the superheavy gauge boson exchanges. The proton lifetime via dimension six operators, with $M \sim 2 \times 10^{16}$ GeV, is obtained around $10^{36}$ years \cite{Ellis:2002vk,Li:2009fq,Li:2010dp,Rehman:2018nsn} for the dominant channel, $p \rightarrow e^+ \pi^0$. Similar to $SU(5)$ model, the prediction of proton lifetime with dimension six operators also lies beyond the scope of Hyper-Kamiokande \cite{HKTDR}.  For a recent comparison of proton decay rates for flipped versus unflipped $SU(5)$ models see \cite{Ellis:2020qad}.

\section{Gauge coupling unification in R-symmetric SU(5)}
An important issue regarding the appearance of massless exotics in an $R$-symmetric $SU(5)$ model of inflation \cite{Kyae:2004ft} is briefly discussed here. Specifically, the un-eaten modes of the adjoint Higgs superfield, that is the octet and the triplet components of $24_H$, do not acquire masses due to $R$-symmetry. This is also discussed in shifted and smooth versions of SUSY hybrid inflation models of $SU(5)$ \cite{Khalil:2010cp,Rehman:2014rpa}. In general, for GUTs based on a simple gauge group this issue was first highlighted in \cite{Barr:2005xya} and then further elaborated in \cite{Fallbacher:2011xg}. These massless particles later acquire TeV scale masses due to soft SUSY breaking effects and consequently ruin the gauge coupling unification, an attractive feature of MSSM. Note that the gauge coupling unification is not predicted in $FSU(5)$ model even though the above issue of massless states does not arise in this model.

To circumvent this issue, vectorlike particles of TeV scale mass were added in \cite{Khalil:2010cp} to achieve gauge coupling unification. But these vectorlike particles do not form a complete $SU(5)$ irreducible representation. Therefore, we follow the trick employed in \cite{Masoud:2019gxx} which is described below in detail for our model. 

\begin{figure}[ht]
	\centering
	\includegraphics[width=0.7\linewidth]{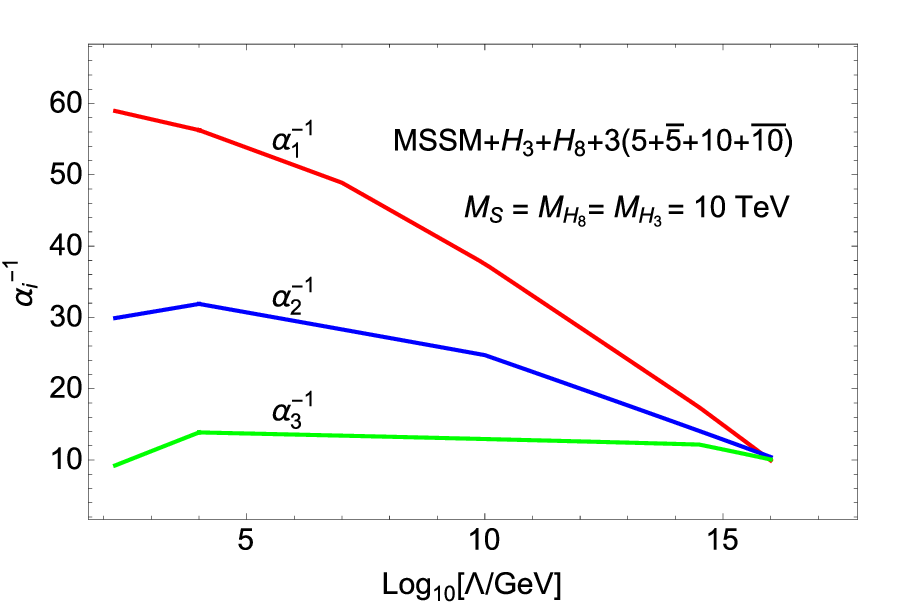}
	\caption{The gauge coupling unification in $SU(5)$ with three additional vectorlike families, $5 \oplus \overline{5}\oplus 10 \oplus \overline{10}$. The effective SUSY breaking scale, $M_S$, along with the octet, $H_3$ and the triplet, $H_8$, components of $24_H$ are taken at 10 TeV. The masses of vectorlike components are chosen as, $M_{Q\oplus\overline{Q}}= M_{U\oplus\overline{U}}= 10^{16}$ GeV, $M_{E\oplus\overline{E}}=10^7$ GeV, $M_{D\oplus\overline{D}}=3\times 10^{14}$ GeV and $M_{L\oplus\overline{L}}=10^{10}$ GeV.}
\label{fig:GCUSU5families3}
\end{figure}

Following \cite{Masoud:2019gxx}, we can add multiple vectorlike families, $5\oplus 10\oplus \overline{5}\oplus \overline{10}$. In the absence of TeV scale exotics, these families keep the gauge coupling unification feature of MSSM, as shown explicitly in \cite{Babu:2008ge}. However, in the presence of TeV scale exotics with additional vectorlike families the gauge coupling unification is destroyed unless we assume appropriate mass splitting among the components of $SU(5)$ multiplets. This is achieved in a similar way we solve the doublet-triplet problem. See \cite{Masoud:2019gxx} for more detail of generating mass splitting. Note that the fine-tunning required in generating the mass splitting of vectorlike components is usually far less than is needed in the doublet-triplet splitting. For the model under consideration, we take the components, $L\oplus \overline{L}$ in $5\oplus \overline{5}$ and $E \oplus\overline{E}$ in $10\oplus\overline{10}$, light while all other remaining components, $D\oplus\overline{D}$ in $5\oplus \overline{5}$ and $Q\oplus\overline{Q}\oplus U\oplus\overline{U}$ in $10\oplus\overline{10}$, are heavy. A typical scenario with three vectorlike families is depicted in Fig.~\ref{fig:GCUSU5families3}. Here, all families are treated on equal footing i.e. the mass splitting is the same for all families. The specific mass values of the various components and scales are mentioned in the caption of Fig.~\ref{fig:GCUSU5families3}. In particular, we set the SUSY breaking scale, $M_S$, equal to 10 TeV in order to avoid dimension five fast proton decay discussed in the previous section. It is interesting to note that the gauge coupling unification scale can be accommodated around $10^{16}$ GeV which was assumed in the earlier sections for making predictions of the various inflationary observables.

\section{\large{\bf Summary}}
The supersymmetric realization of new inflation is discussed for GUTs based on $SU(5)$ and $FSU(5) \equiv SU(5)\times U(1)$ gauge groups. As introduced in \cite{Senoguz:2004ky}, this class of new inflation models basically utilizes the framework of $R$-symmetric SUSY hybrid inflation model with  additional $Z_n$ symmetry. The SM gauge singlet component of GUT Higgs field plays the role of inflaton in these models. The monopole problem of $SU(5)$ model is naturally resolved. The realization of light neutrino masses and the reheating  with leptogenesis both are easily incorporated in the minimal $SU(5)$ model. The issue of gauge coupling unification with the appearance of massless fields in $R$-symmetric $SU(5)$ model is tackled with additional vectorlike families. The doublet-triplet splitting is naturally achieved in $FSU(5)$ model whereas a fine-tuned solution is assumed in case of $SU(5)$ model. The key role played by the $U(1)_R \times Z_n$ symmetry in suppressing the various fast proton decay operators is highlighted for both models. The proton decay is found to proceed via dimension six operators with a lifetime of order  $10^{36}$ years for the dominant decay channel, $p \rightarrow e^+ \pi^0$. A related issue of generating light neutrino masses and the reheating with successful leptogenesis in $FSU(5)$ is also discussed with possible solutions. The predictions of the various inflationary parameters are compared with the recent Planck data bounds on the scalar spectral index, $n_s$. The predicted ranges of the tensor to scalar ratio, $10^{-12} \lesssim r \lesssim 10^{-8}$, the running of the scalar spectral index, $7 \times 10^{-10} \lesssim |dn_s/d\ln k| \lesssim 10^{-3}$, and the reheat temperature, $3 \times 10^{7}\text{ GeV }\lesssim T_r \lesssim 2 \times 10^{9}$ GeV, are found within Planck data bounds on $n_s$.

\section*{Acknowledgments}
We thank Qaisar Sahfi for useful discussion.


\end{document}